\documentclass[reprint,superscriptaddress,amsmath,amssymb,aps,prx,longbibliography]{revtex4-2}
\usepackage{amsmath}
\usepackage{graphicx}
\usepackage{longtable}
\usepackage{colortbl}
\usepackage[english]{babel}
\usepackage{array}
\usepackage{booktabs}
\usepackage[colorlinks=true,linkcolor=blue,anchorcolor=blue,filecolor=blue,urlcolor=blue,citecolor=blue]{hyperref}
\graphicspath{{graphs/}}

\DeclareUnicodeCharacter{2212}{-}
\begin{document}

\title{Many-Body Effects in the X-ray Absorption Spectra of Liquid Water}
\author{Fujie Tang} \affiliation{Department of Physics, Temple University, Philadelphia, Pennsylvania 19122, USA}
\author{Zhenglu Li} \affiliation{Department of Physics, University of California at Berkeley, Berkeley, California 94720, USA}
\affiliation{Materials Sciences Division, Lawrence Berkeley National Laboratory, Berkeley, California 94720, USA}
\author{Chunyi Zhang} \affiliation{Department of Physics, Temple University, Philadelphia, Pennsylvania 19122, USA}
\author{Steven G. Louie} \affiliation{Department of Physics, University of California at Berkeley, Berkeley, California 94720, USA}
\affiliation{Materials Sciences Division, Lawrence Berkeley National Laboratory, Berkeley, California 94720, USA}
\author{Roberto Car}
\thanks{Corresponding author. Email: rcar@princeton.edu}
\affiliation{Department of Chemistry, Princeton University, Princeton, New Jersey 08544, USA}

\author{Diana Y. Qiu}
\thanks{Corresponding author. Email: diana.qiu@yale.edu}
\affiliation{Department of Mechanical Engineering and Materials Science, Yale University, New Haven, Connecticut 06520, USA}
\author{Xifan Wu}
\thanks{Corresponding author. Email: xifanwu@temple.edu}
\affiliation{Department of Physics, Temple University, Philadelphia, Pennsylvania 19122, USA}
\date{\today}

\begin{abstract}
X-ray absorption spectroscopy (XAS) is a powerful experimental technique to probe the local order in materials with core electron excitations. Experimental interpretation requires supporting theoretical calculations. For water, these calculations are very demanding and, to date, could only be done with major approximations that limited the accuracy of the calculated spectra. This prompted an intense debate on whether a substantial revision of the standard picture of tetrahedrally bonded water was necessary to improve the agreement of theory and experiment. Here, we report a new first-principles calculation of the XAS of water that avoids the approximations of prior work thanks to recent advances in electron excitation theory. The calculated XAS spectra, and their variation with changes of temperature and/or with isotope substitution, are in excellent quantitative agreement with experiments. The approach requires accurate quasi-particle wavefunctions beyond density functional theory approximations, accounts for the dynamics of quasi-particles and includes dynamic screening as well as renormalization effects due to the continuum of valence-level excitations. The three features observed in the experimental spectra are unambiguously attributed to excitonic effects. The pre-edge feature is associated to a bound intramolecular exciton, the main-edge feature is associated to an exciton localized within the coordination shell of the excited molecule, while the post-edge one is delocalized over more distant neighbors, as expected for a resonant state. The three features probe the local order at short, intermediate, and longer range relative to the excited molecule. The calculated spectra are fully consistent with a standard tetrahedral picture of water.

 \end{abstract}

\maketitle

\section{Introduction}
\par Water is the most important material on earth. Its structure is defined by a hydrogen bond (H-bond) network whose organization gives rise to its characteristic properties, such as an increased density upon melting, decreased viscosity under pressure, a density maximum at 4 $\rm ^\circ C $, high surface tension, and many more~\cite{eisenberg1969,Stillinger451}. Over decades, numerous advanced experimental techniques have been applied to water in order to reveal the precise character and arrangement of the H-bond network~\cite{Soper2000,wernet2004,Winter2004,tse2008,Bakker2010,Perakis2016,fransson2016x,Saykally2017}. Among them, the core-level X-ray absorption spectroscopy (XAS) has emerged as a powerful local probe of the water structure~\cite{wernet2004,tse2008,wernet2014,wernet2016,Saykally2017,Carbone2020}, which is complementary to the averaged structural information obtained in other scattering experiments. First-principles calculations are indispensable for an unambiguous interpretation of the underlying water structure from the measured spectra~\cite{hetenyi2004,cavalleri2004x,Prendergast2005,Chen2010,vinson2012,kong2012,fransson2016p,sun2017x,sun2018,Zhovtobriukh2018,Zhovtobriukh2019}. However, theoretical modeling of XAS in water has proved a challenging subject for the past twenty years. Despite extensive studies, a consensus has yet to be reached~\cite{fransson2016p,fransson2016x,Zhovtobriukh2018,Zhovtobriukh2019}.

\par In XAS experiment, a negatively-charged core electron is excited by absorbing a high-energy photon, and leaves behind a positively-charged core-hole with which it can interact to form an exciton~\cite{PARRATT1959,Rehr2000}. The electron-hole interaction is expected to be relatively strong in water, which has a small dielectric constant at characteristic electronic frequencies due to its large band-gap, and the attraction is further strengthened by the disordered liquid structure, which facilitates the localization of the excited electron near the core-hole. Consequently, excitonic effects are believed to play an important role in the XAS of water~\cite{Olovsson2009,vinson2011,vinson2012,Gulans2014,Vorwerk2020,ONO2015,Noguchi2015,Vorwerk2017}. In first-principles calculations, the accurate treatment of many-electron effects in two-particle excitations is computationally formidable~\cite{Onida2002}. Most studies to date adopt the static core-hole approximation~\cite{hetenyi2004,cavalleri2004x,Prendergast2005,Chen2010,kong2012,sun2017x,sun2018,zhangchunyi2020}, which makes the assumption that the hole is described by the static potential of a single core-level atomic orbital, thus simplifying the correlated electron-hole excitation to an equivalent one-particle excitation process. Following this assumption, improved descriptions of the excited electron have been developed and applied in the past decade, where the electron has been treated within an independent electron approximation~\cite{Prendergast2005}, Slater's transition-state method~\cite{Slater1969,slater1970,hetenyi2004,cavalleri2004x}, and self-energy approaches~\cite{Chen2010,kong2012,sun2017x,sun2018,zhangchunyi2020}. Nevertheless, the reported spectra fluctuate among the various methods and discrepancies remain between theory and experiment, raising concerns about the neglect of the core-hole dynamics and resultant renormalization of the core wavefunction in the presence of electron-hole interactions. This effect can be rigorously accounted for within the GW plus Bethe-Salpeter equation (GW-BSE) method in many-body perturbation theory in which the excitations are correctly described as coherent superpositions of band electron-core hole pairs~\cite{Hybertsen1986,rohlfing2000,Onida2002,Olovsson2009,vinson2011,Gulans2014,Vorwerk2017}. Because of the computational complexity, the full GW-BSE approach has rarely been applied to study the XAS of water. A first attempt was made by Vinson et al.~\cite{vinson2012} a few years ago, which showed reasonable agreement with experiment. However, the reported spectrum~\cite{vinson2012} lacks a discernible post-edge feature and has a narrower spectral width than experiment.

\par The current difficulty in reconciling theoretical and experimental spectra reflects long-standing challenges in the theoretical modeling of water, both for the electronic ground state and the excited states. For the ground state, density functional theory (DFT) tends to overestimate the H-bonding strength~\cite{Kuhne2009,DiStasio2014,Chen2017}. The H-bonds form due to a combination of electrostatic attraction and covalency, which originates from the charge transfer between a H-bond donor and acceptor. This charge transfer is greatly overestimated in DFT due to the spurious self-interaction and missing derivative discontinuity of commonly used exchange-correlation (XC) potentials~\cite{cohen2008i,cohen2012c}. This error is carried over into the calculation of the excited states in conventional GW-BSE calculations, which approximate the quasiparticle (QP) wavefunctions as the DFT orbitals~\cite{Hybertsen1986,rohlfing2000,Onida2002}. One way to resolve this is to self-consistently determine the Green's function in the GW approach to obtain the QP wavefunctions~\cite{caruso2012,faber2013,Kaplan2015}. Additionally, XAS processes in water are dominated by transitions from the oxygen core-hole state to conduction band states near the Fermi level. At the same photon energies, electrons may also be excited from valence states to high-energy continuum states. Both processes (which couple quantum mechanically to determine the excitation spectrum) are included in experiments~\cite{PARRATT1959}. However, all theoretical studies thus far have neglected the second process involving the valence to continuum transitions~\cite{hetenyi2004,cavalleri2004x,Prendergast2005,Chen2010,vinson2012,kong2012,fransson2016p,sun2017x,sun2018,Zhovtobriukh2018,Zhovtobriukh2019}. A complete treatment of both processes requires one to solve the GW-BSE over the entire Hilbert space of transitions between occupied and unoccupied states consistent with the energy of the photon. In water, this represents a basis of more than 860,000 single-particle states for a box of 32 water molecules. Due to the computational burden, such an effect has yet to be examined in water.

\par Here, we develop an efficient GW-BSE workflow that allows us to calculate the XAS of liquid water from \textit{ab initio} with unprecedented accuracy, by combining state-of-the-art methods for both the ground-state---through the use of path-integral deep potential molecular dynamics (PI-DPMD)~\cite{zhang2021m} for the atomic structure---and excited states---through the use of QP wavefuctions calculated within the static GW approximation, frequency-dependent and fully non-local treatment of the dielectric response, and a Hilbert space downfolding approach newly developed to account for the coupling of the core- and valence-level transitions from the electron-hole kernel of the BSE~\cite{benedict2002,diana2021,DianaTBD}. We find that a combination of all these techniques allows us to reproduce the experimental XAS to a high level of accuracy for both the relative energies of the pre-, main- and post-edge features and their spectral line shape, and their physical origins. In particular, we find that the use of self-consistent QP wavefunctions is especially important, since the wavefunction renormalization due to the electron self energy promotes intramolecular excitations over intermolecular excitations by reducing the degree of charge transfer in the H-bonds. This, in turn, shifts the oscillator strength from the main-edge to the pre-edge, improving agreement with experiment. Our calculations accurately predict not only the XAS of water but also the subtle changes to the spectrum under elevated temperature and isotope substitution. Our work solves long-existing challenges in the $ab$ $initio$ prediction of the XAS of water, establishing the essential role of electron-hole and many-electron interactions in the prediction of the spectral features and thus drawing an unambiguous path between the excitation spectrum and the underlying molecular structure. The physical understanding and computational approach developed here will be applied in the future to understand XAS experiments in aqueous solutions, confined water, and water at interfaces.

\section{Theoretical Methods}
\subsection{Molecular Dynamics Simulation}
\par The configurations of liquid water used for the X-ray absorption spectra (XAS) calculations were generated by path-integral deep potential molecular dynamics (PI-DPMD)~\cite{linfeng2018} using the deep potential model reported in our previous work~\cite{zhang2021m}. In particular, the deep potential model was trained on the density functional theory (DFT) data obtained with the hybrid strongly constrained and appropriately normed (SCAN0) functional~\cite{zhang2021m}. The cubic cells contain 32 water molecules while the size of the supercell for each simulation was adjusted to have the same density in experiment. Our PI-DPMD simulations of H$_2$O were performed in the {\it NVT} ensemble at both {\it T} = 300 K and 330 K using periodic boundary conditions with the cell sizes fixed at 9.708 $\rm \AA$ and 9.724 $\rm \AA$, respectively. To study isotopic effects on the XAS, we also performed a PI-DPMD simulation of D$_2$O at 300 K with the cubic cell size was fixed at 9.708 $\rm \AA$. In PI-DPMD, the Feynman paths were represented by 8-bead ring polymers coupled to a color noise generalized Langevin equation thermostat (i.e. PIGLET)~\cite{Ceriotti2009}. All the PI-DPMD simulations used well equilibrated $\sim$500 ps long trajectories.

To compute the XAS spectra of liquid water, we need to average over snapshots of the molecular dynamics trajectories. However, due to the huge computational cost of the GW-BSE calculation, we could only use a limited number of snapshots to calculate the spectra. In order to select the most representative snapshots, we assigned to each snapshot $i$ a score function $f(i)$ that measures the deviation of the structure at snapshot $i$ 
relative to the average structure in a trajectory.  The score function $f(i)$ is defined as:
	\vspace{-0.5em}
\begin{equation}
		f(i)=\sum_{k=1}^{7}\left| X^i_k-\bar{X}_k\right|/|\bar{X}_k|.
\end{equation}
Here $X^i_k$ $(k=1,\ldots,7)$ are descriptors of intra-molecular structure, structure of H-bond network, and thermodynamic properties at snapshot $i$. Specifically, $X^i_{k=1}$ is the average proton transfer distance $\delta=r(\rm O\cdots H)$ - $r(\rm OH)$~\cite{Lu2014}, $X^i_{k=2}$ is the average covalent bond length $r(\rm OH)$, $X^i_{k=3}$ is the average number of H-bonds, $X^i_{k=4}$ is the average H-bond length, $X^i_{k=5}$ is the average O-O nearest neighbor distance, $X^i_{k=6}$ is the average local structure index~\cite{Biswajit2015}, and $X^i_{k=7}$ is the instantaneous temperature defined by the average kinetic energy of the atoms. The corresponding averages over the entire trajectory are denoted by $\bar{X}_k$. Near the minimum of the score function, we selected two independent snapshots for H$_2$O and D$_2$O at 300 K, respectively and one snapshot for H$_2$O at 330 K. Each snapshot contains 8 different molecular configurations corresponding to the 8 beads representing the Feynman paths. In order to study the response to a temperature change, we also selected a snapshot for H$_2$O at 330 K. We note that the snapshots identified in this way, i.e., by minimal values of $f(i)$, have descriptors $X^i_k$ very close to their mean values $\bar{X}_k$, with deviations not larger than $0.5\%$. This suggests that the selected snapshots can serve as representative water structures in the GW-BSE calculations.

\subsection{The GW-BSE XAS Calculation}

\par  Our GW-BSE XAS calculation was performed using a modified version of the BerkeleyGW~\cite{Hybertsen1986,rohlfing2000,Deslippe2012} package.  To obtain the QP wavefunctions, we first calculated the mean-field wavefunctions as the starting point for our GW-BSE calculation using the DFT at the level of the generalized gradient approximations (GGA) of Perdew, Burke and Ernzerhof (PBE)~\cite{perdew1996}, as implemented in Quantum ESPRESSO~\cite{Giannozzi2017}. The multiple-projector norm-conserving pseudopotentials that match the all electron potentials for oxygen and hydrogen were generated by using the ONCVPSP package~\cite{hamann2013}. The DFT wavefunction was set to have a 200 Ry plane wave cutoff to converge the description of the core electrons. Our GW-BSE calculation was conducted by using a modified version of the BerkeleyGW~\cite{Hybertsen1986,rohlfing2000,Deslippe2012} package. The GW calculation was done only at the $\Gamma$ point and we used a 20 Ry cutoff (with 10000 bands included in the sum over empty bands) for the plane-wave components of the dielectric matrix. A G$_1$W$_0$ self-consistent calculation was firstly performed with the static COHSEX approximation in order to improve the quasiparticle (QP) wavefunctions. In this step, for the 32-molecule simulation cell, 160 occupied states and 320 unoccupied states were self-consistently updated according to the off-diagonal matrix elements of the self-energy operator. Since the number of conduction bands involved in the self-consistent updating procedures could affect the spectral shape, we performed a convergence test, the result of which is shown in Fig. S1. The spectrum shows a good convergence of the pre-edge and main-edge, while the post-edge contains some small fluctuations. Due to the heavy computational cost of the self-consistent procedure with a larger number of bands, we chose to use 320 conduction bands and 160 occupied states in the self-consistent update of the QP wavefunction. Then, a standard one-shot G$_0$W$_0$ calculation was performed with 10000 bands (where the 160 occupied and the first 320 unoccupied bands are from self-consistent COHSEX), and the frequency dependence in the dielectric matrix is captured using the Hybertsen-Louie generalized plasmon-pole model (HL-GPP)~\cite{Hybertsen1986}. The BSE calculation was done with 32 core states and 160 unoccupied states, which are enough to cover the energy range in which we are interested. For all the calculations, we solved the electron-hole excitations within the GW-BSE approach in the Tamm-Dancoff approximation (TDA)~\cite{zagoskin1998,rohlfing2000}. We calculated the spectral XAS intensity using Eq. 26 in Ref.~\cite{rohlfing2000}, shown as below:
	\vspace{-0.5em}
\begin{equation}\label{eqn:2}
\epsilon_{2}(\omega)=\frac{16{\pi}^2 e^2}{\omega^2}\sum_{S}\left|{\boldsymbol e}\cdot\left<0\left|\boldsymbol{v}\right|S\right>\right|^2\delta(\omega-\Omega_S)  
	\vspace{-0.5em}
\end{equation}

The optical transition matrix elements are given by:
	\vspace{-0.5em}
\begin{equation}
	\left<0\left|\boldsymbol{v}\right|S\right>=\sum_v^{\rm hole}\sum_c^{\rm elec}\sum_{\boldsymbol{k}}A_{vc\boldsymbol{k}}^{S}<v\boldsymbol{k}|\boldsymbol{v}|c\boldsymbol{k}>
	\vspace{-0.5em}
\end{equation}
where the $v$($c$) denotes core-valence (conduction) states, represented by the DFT eigenstates or QP eigenstates in the G$_0$W$_0$-BSE@PBE and G$_0$W$_0$-BSE@sc-G$^{\rm static}$W$_0$ approaches, respectively; $\boldsymbol{e}$ is the direction of the polarization of light. 
\par In order to calculate the optical transition matrix, the matrix element $<v\boldsymbol{k}\left|\boldsymbol{v}\right|c\boldsymbol{k}>$ needs to be evaluated for the $\boldsymbol{k}$-point mesh that is used to sample the Brillouin zone. Under periodic boundary condition, the velocity operator $\boldsymbol{v}$ is determined by the commutator of the Hamiltonian $H$ and the position operator $\boldsymbol{r}$ as: $\boldsymbol{v} = i[H, \boldsymbol{r}]$. Therefore, the matrix element $<v\boldsymbol{k}\left|\boldsymbol{v}\right|c\boldsymbol{k}>$ can be exactly computed following Ref.~\cite{Deslippe2012} as:
\begin{align}\label{eqn:news1}
	<v\boldsymbol{k}\left|\boldsymbol{v}\right|c\boldsymbol{k}>&=<v\boldsymbol{k}\left|i[H, \boldsymbol{r}]\right|c\boldsymbol{k}>\\\nonumber
	&=i(E_v-E_c)<v\boldsymbol{k}|\boldsymbol{r}|c\boldsymbol{k}>\\\nonumber
	&\simeq i(E_v-E_c)\lim_{\boldsymbol{q}\rightarrow0}\frac{<v\boldsymbol{k}+\boldsymbol{q}|e^{i\boldsymbol{q}\cdot\boldsymbol{r}}-1|c\boldsymbol{k}>}{i\boldsymbol{q}}\\\nonumber
	&=(E_v-E_c)\lim_{\boldsymbol{q}\rightarrow0}\frac{<v\boldsymbol{k}+\boldsymbol{q}|e^{i\boldsymbol{q}\cdot\boldsymbol{r}}|c\boldsymbol{k}>}{\boldsymbol{q}}.
\end{align}
In practice, the $<v\boldsymbol{k}\left|\boldsymbol{v}\right|c\boldsymbol{k}>$ is numerically computed by evaluating the limit via finite difference method for a small value of $\boldsymbol{q}$ along the direction of the polarization of light as shown in~\eqref{eqn:news1}. As described above, the calculation of matrix element $<v\boldsymbol{k}\left|\boldsymbol{v}\right|c\boldsymbol{k}>$ by the velocity operator $\boldsymbol{v}$ demands the additional calculation of the QP wavefunctions for $\boldsymbol{k}$ shifted by $\boldsymbol{q}$. This is a severe computational burden when the QP wavefunctions are obtained by self-consistently determining the Green's function in the G$_0$W$_0$-BSE@sc-G$^{\rm static}$W$_0$ approach. In order to alleviate such computational burden, one may approximate the velocity operator $\boldsymbol{v}$ by the momentum operator $\boldsymbol{p}=-i\nabla$ as suggested in Ref.~\cite{Deslippe2012}, i.e.: 
\begin{equation}\label{eqn:news2}
	<v\boldsymbol{k}\left|\boldsymbol{v}\right|c\boldsymbol{k}>=<v\boldsymbol{k}\left|\boldsymbol{p}+i[V_{\rm ps},\boldsymbol{r}]\right|c\boldsymbol{k}>\simeq <v\boldsymbol{k}\left|\boldsymbol{p}\right|c\boldsymbol{k}>.
\end{equation}
As shown in~\eqref{eqn:news2}, the approximated calculation via the momentum operator $\boldsymbol{p}$ is equivalent to neglecting the commutator between $\boldsymbol{r}$ and the non-local part of the pseudopotentail $V_{\rm ps}$~\cite{rohlfing2000}. In our GW-BSE calculation, we approximated the velocity operator $\boldsymbol{v}$ by the momentum operator $\boldsymbol{p}$ to calculate the optical transition matrix. We compare in Fig. S2 the XAS spectra calculated for one structure from an equilibrated trajectory (PI-DPMD of H$_2$O at 300 K) using the G$_0$W$_0$-BSE@PBE approach with the velocity operator $\boldsymbol{v}$ and the momentum operator $\boldsymbol{p}$, respectively. As one can see, with the contributions from the nonlocal parts of the pseudopotentials, the G$_0$W$_0$-BSE@PBE XAS spectrum shows a higher pre-edge and main-edge peaks, as well as a reduced post-edge peak, suggesting that improving the calculation of the transition matrix elements could further improve the agreement of the calculation with experiment. Therefore, it will be interesting to compute the G$_0$W$_0$-BSE@sc-G$^{\rm static}$W$_0$ XAS spectrum using the velocity operator $\boldsymbol{v}$ in future studies.


\subsection{The Numeric Implementations of the $S$ Approximation}

\par We argued that when solving the BSE in a restricted subspace $\cal{A}$ (transitions from core states to the low-lying conduction band states) because of the inaccessible calculation in the full Hilbert space ($\cal{A}\oplus \cal{B}$, where subspace $\cal{B}$ includes the transitions between valence band states and continuum conduction band states), the exchange term $K^{x}_{\rm eff}$ should be replaced with an effective exchange term which is reduced (screened) with respect to the usual exchange term $K^{x}$ according to Benedict's $S$ approximation~\cite{benedict2002}. Here we show how to modify the BSE kernel to implement the $S$ approximation by screening the exchange term $K^{x}_{\rm eff}$~\cite{diana2021}. The modified BSE kernel can be decomposed by $K^{\rm eh}=K^{d}+K_{\rm eff}^{x}$, where the $K^{x}_{\rm eff}$ is defined as:
	\vspace{-1.0em}
\begin{widetext}
\begin{equation}\label{eqn:4}
		\vspace{-1.0em}
	\left<vc{\boldsymbol{k}}\left|{{K}^{x}_{\rm eff}}\right|{v'c'}\boldsymbol{k}'\right>=\sum_{GG'}{M}^*_{vc\boldsymbol{k}}({\boldsymbol{Q}},{\boldsymbol{G}})\overline{W}_{GG'}{(\omega={\rm 0},{\boldsymbol{Q}})}{M}_{v'c'\boldsymbol{k}'}(\rm{\boldsymbol{Q},{\boldsymbol{G}}}').   
\end{equation}
\end{widetext}
where $M_{vc\boldsymbol{k}}({\boldsymbol{Q}},{\boldsymbol{G}})=\left<{v}{\boldsymbol{k}}+{\boldsymbol{Q}}\left|{e}^{{i}({\boldsymbol{Q}}+{\boldsymbol{G}})*{\boldsymbol{r}}}\right|{c}{\boldsymbol{k}}\right>$, ${\boldsymbol{G}}$ is a reciprocal lattice vector and ${\boldsymbol{Q}}$ is the center of mass momentum of the electron-hole pair. By using the $S$ approximation, the usual bare Coulomb interaction, $\it{v}({\boldsymbol{Q}+\boldsymbol{G}})\delta_{GG'}$, is replaced by $\overline{W}_{GG'} (\omega=0, \boldsymbol{Q})$, which is the screened Coulomb interaction in the static limit. The new screened Coulomb term $\overline{W}$ could be constructed by using the polarizability of subspace $\cal{B}$ created by the positive electron-hole excitations~\cite{diana2021} within the random-phase approximation (RPA).  
	\vspace{-0.5em}
\begin{equation}\label{eqn:5}
\overline{W}_{GG'} (\omega=0,\boldsymbol{Q})=\overline{\epsilon}_{GG'}^{-1}(\omega=0,{\boldsymbol{Q}})\it{v}({\boldsymbol{Q}}+{\boldsymbol{G}}),
\end{equation}
where $\overline{\epsilon}$ is a dielectric matrix of the form,
		\vspace{-1.0em}
\begin{equation}\label{eqn:6}
		\vspace{-0.em}
\overline{\epsilon}_{GG'}(\omega=0,{\boldsymbol{Q}}) = \delta_{GG'} - \it{v}({\boldsymbol{Q}}+{\boldsymbol{G}})\frac{1}{2}\overline{\chi}^{\rm 0}_{GG'}(\omega = {\rm 0}, {\boldsymbol{Q}}). 
\end{equation}
The factor of $\frac{1}{2}$ excludes the electron-hole pairs that backward propagating time. The $\overline{\chi}^{0}_{\rm GG'}$ is the static, noninteracting RPA polarizability due to all electron-hole pairs not in the subspace $\cal{A}$, which could be evaluated by using the formula:
	\vspace{-1.0em}
\begin{widetext}
\begin{equation}\label{eqn:7}
			\vspace{-1.0em}
\overline{\chi}^{0}_{GG'}({\boldsymbol{Q}})=\sum_{n\in{A+B}}^{\rm{occ}}\sum_{n'\in{A+B}}^{\rm{unocc}}\sum_{{\boldsymbol{k}}}\frac{{M}^*_{nn'{\boldsymbol{k}}}({\boldsymbol{Q}},{\boldsymbol{G}}){M}_{nn'{\boldsymbol{k}}}({\boldsymbol{Q}},{\boldsymbol{G}}')}{{E}_{n{\boldsymbol{k}}+{\boldsymbol{Q}}}-{E}_{n'{\boldsymbol{k}}}}-\left[\chi^{\rm AA}\right]^{0}_{GG'}({\boldsymbol{Q}}),
\end{equation}
\end{widetext}
where the $\left[\chi^{\rm AA}\right]^{0}$ is the noninteracting RPA polarizability formed the transitions from the oxygen core states to the low-lying conduction band states, which belongs to the subspace $\cal{A}$,
	\vspace{-1em}
\begin{equation}\label{eqn:8}
\left[\chi^{\rm AA}\right]^{0}_{GG'}({\boldsymbol{Q}})=\sum_{n\in{A}}^{\rm occ}\sum_{n'\in{A}}^{\rm unocc}\sum_{{\boldsymbol{k}}}\frac{{M}^*_{nn'{\boldsymbol{k}}}({\boldsymbol{Q}},{\boldsymbol{G}}){M}_{nn'{\boldsymbol{k}}}({\boldsymbol{Q}},{\boldsymbol{G}}')}{{E}_{n{\boldsymbol{k}}+{\boldsymbol{Q}}}-{E}_{n'{\boldsymbol{k}}}}.  
\end{equation}
\par Based on Eqs. (\ref{eqn:4}-\ref{eqn:8})~\cite{diana2021}, we have numerically determined the effective exchange term ($K^{x}_{\rm eff}$) in the GW-BSE kernel in~\eqref{eqn:4} for one structure from an equilibrated trajectory (PI-DPMD of H$_2$O at 300 K). The subspace $\cal{A}$ was constructed by the electron-hole excitations from 1s core states of oxygen atoms to 160 conduction states, and the subspace $\cal{B}$ was spanned by the electron-hole excitations from valence bands to the continuum states up to 500 eV. The resulting XAS spectra based on the $S$-approximation is shown by dashed lines are Fig. S3 (a), (b) and (c) for the incoming light polarized along X, Y and Z directions, respectively. The RPA calculation of the polarizability of the subspace $\cal{B}$, outlined above, is very expensive. Fortunately, neglecting the $\boldsymbol{Q}$ dependence of the effective dielectric matrix in~\eqref{eqn:5} turns out to be an excellent approximation. Setting
${\bar{\epsilon}}^{-1}_{GG’} (\boldsymbol{Q}) = \frac{1}{\bar{\epsilon}}\delta_{GG’}$, where $\bar{\epsilon}$ is a constant, leads to $K^x_{\rm eff} = K^x / {\bar{\epsilon}} = \alpha K^x$. Here we defined an effective screening parameter $\alpha = 1 / {\bar{\epsilon}}$. By varying $\alpha$ from 0 to 1 in increments of 0.1, we found that $\alpha = 0.8$ yields nearly identical spectra to those computed using the full RPA expression as shown in Fig. S3. We further checked that the optimal alpha value is rather insensitive to the molecular structure and works equally well for H$_2$O at 330 K and D$_2$O at 300 K. Thus, we used $\alpha=0.8$ in the spectral calculations reported in the main text. We remark that setting $\alpha=0.8$ corresponds to an effective dielectric constant $\bar{\epsilon} = 1.25$, which is smaller than the electronic dielectric constant of water ($\epsilon_{\infty} = 1.8$). This is consistent with the $S$-approximation~\cite{Benedict2003} according to which $1 < \bar{\epsilon} < \epsilon_{\infty}$, and the lower limit ($\bar{\epsilon} = 1$) is found when the dimension of the subspace $\cal{B}$ becomes infinite.

\section{Theoretical XAS of Water}

\par
\begin{figure}[htbp]
	\setlength{\abovecaptionskip}{0.cm}
	\includegraphics[width=3.4in]{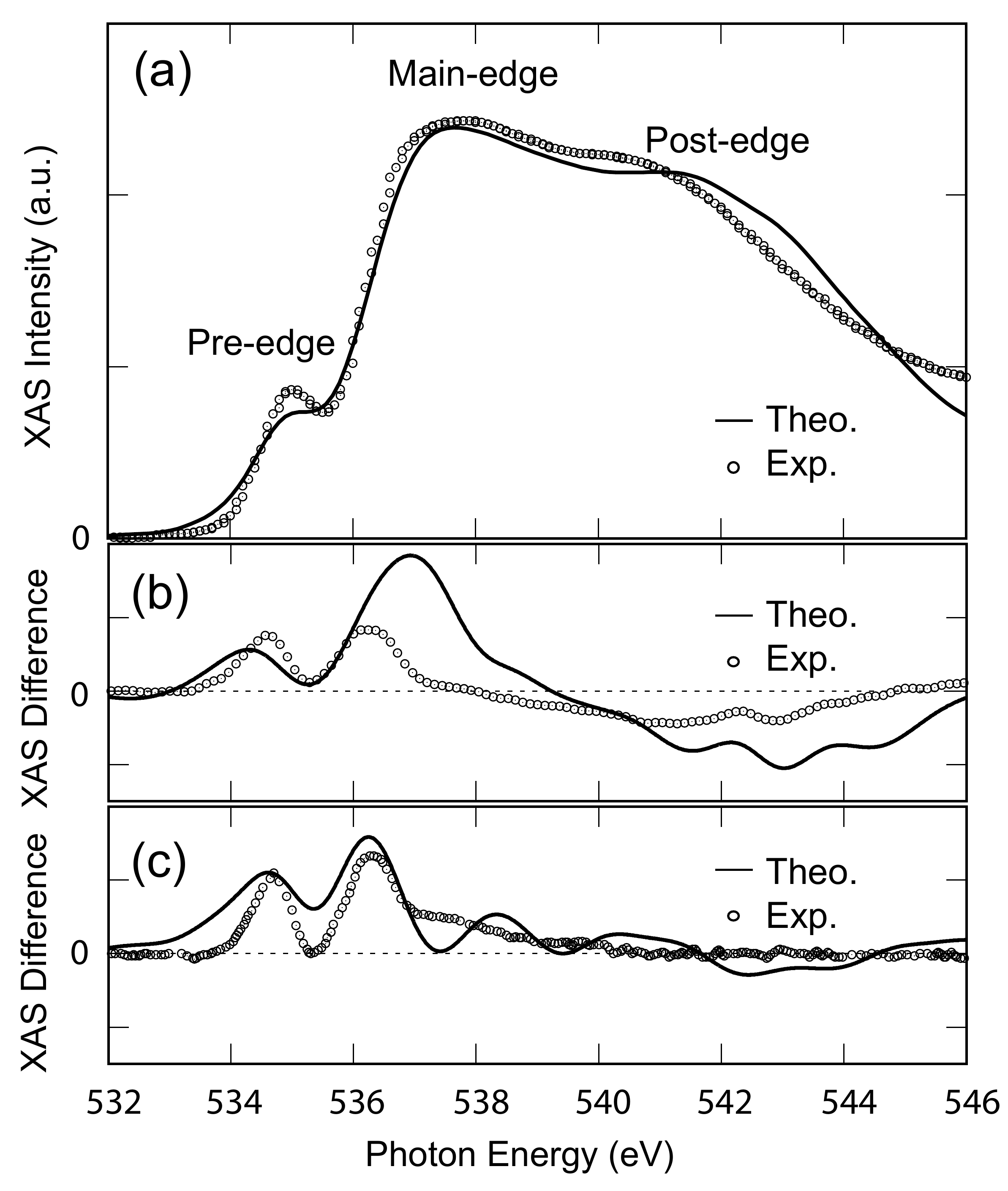}
	\caption{\label{fig:xas_overall}
		(a) The experimental (circles) and theoretical (solid) oxygen K-edge XAS of liquid water. The theoretical XAS was generated at the GW-BSE level using the atomic configuration from a PI-DPMD simulation at 300 K. The experimental data is taken at 296 K from Ref.~\cite{wernet2014}. The theoretical spectrum is rigidly shifted by $\sim$16 eV to align with the pre-edge peak of the experimental spectrum. (b) Temperature differential spectra: theory (solid) and experiment~\cite{wernet2014} (circles). The temperature differences are set to $\Delta T = 30$ K. (c) Isotopic differential (H$_2$O-D$_2$O) spectra: theory (solid) and experiment~\cite{wernet2016} (circles).}
	\vspace{-1 em}
\end{figure}

 As shown in Fig.~\ref{fig:xas_overall}(a), we present our theoretical XAS spectrum of liquid water at room temperature. Good agreement can be seen between theory and experiment in both the spectral width and spectral intensities in Fig.~\ref{fig:xas_overall}(a). In particular, the experimentally observed pre-edge, main-edge, and post-edge features at $\sim$535 eV, $\sim$538 eV, and $\sim$541 eV, respectively, are all well reproduced by the rigorous treatment of electron-hole dynamics. We note that the absolute energies in our calculated spectra are red-shifted by $\sim$16 eV compared to the experimental spectra (the onset of the theoretical spectrum is 516 eV compared to 532 eV in experiment). This $\sim$3\% discrepancy in the absolute excitation energy is within the expected margin of error of the GW approach for QP energies in this range. The absolute error of the BSE calculation of exciton binding energies is much smaller, as reflected in the good agreement of the calculated relative peak energies with experiment, since the relevant energy scale is that of the core exciton binding energy on the order of a few eV. To facilitate comparison of features of the calculated and experimental spectra, we aligned all calculated XAS spectra obtained from different individual structures by using the pre-edge peaks of the corresponding experimental spectra. To analyze the character of the electron-hole excitations, we plot the exciton wavefunctions for states at various spectral edges (see Fig. S5 in the Supporting Information). The electron-hole excitations at the pre-edge have strong intramolecular character. The excited electron is primarily localized on the same molecule as the core-hole [as shown in Fig. S5 in Supporting Information], so the exciton can be loosely categorized as Frenkel-like. Compared to the pre-edge, the excited electrons at the main-edge are generally more delocalized. Nevertheless, a certain degree of localization can still be identified. At lower excitation energies near the main-edge, the excited electron is distributed on both the water molecule where the core-hole is localized and the H-bonded water molecules in the coordination shell. Therefore, the main edge is primarily composed of intermolecular electron-hole excitations. In sharp contrast, the electron-hole pairs at post-edge are completely delocalized over the cell, which are consistent with resonant exciton states or interband transitions. 

\par Due to the distinct types of excitations associated with each feature, the pre-edge, main-edge, and post-edge of the XAS spectrum convey short-range, intermediate-range, and long-range structural information about the H-bond network of water~\cite{fransson2016x,Saykally2017}. Thus, XAS can also probe subtle changes in the H-bond network of liquid water due to isotopic and temperature changes. To confirm that our theory captures these changes, we calculate the differential XAS of water under elevated temperature and isotope substitution as presented in Fig.~\ref{fig:xas_overall}(b) and Fig.~\ref{fig:xas_overall}(c), respectively. There is excellent agreement betwen theory and experiment for both the differential temperature and differential isotopic spectra. As the temperature increases, the coordination of water is further distorted away from an ideal tetrahedron by weakened H-bonds, which in turn promotes the localization of excitons on the excited water molecule, making both the pre-edge and main-edge feature more prominent. At the same time, the reduced post-edge in Fig.~\ref{fig:xas_overall}(b) is consistent with the loss of long-range ordering in the H-bond network at higher temperatures, which is a well-known XAS signature when ice melts into water. On the other hand, the spectral difference between heavy and light water is due to nuclear quantum effects, under which hydrogen explores the configuration space more extensively than the heavier deuterium. As shown in Fig.~\ref{fig:xas_overall}(c), more pronounced pre-edge and main-edge features are observed in the differential spectrum between H$_2$O and D$_2$O. While the enhancement of these features is similar to the enhancement seen at elevated temperature, the nuclear quantum effects do not have an identical physical origin to the spectral changes due to elevated temperature. The isotopic changes are due to the higher degree of local disorder of the protons in the light water, both intramolecularly and intermolecularly, which breaks the local tetrahedral coordination and helps localize the electron-hole excitations. The long-range order, as outer shell measured by the oxygen-oxygen distribution functions, is relatively unaffected by this local disorder. Thus, the post-edge, which is a signature of long-range ordering, is much less affected by the isotope substitution in Fig.~\ref{fig:xas_overall}(c). This is consistent with the negligible differences at long-range in the oxygen-oxygen pair distribution functions from neutron scattering experiments~\cite{Soper2008}. The successful reproduction and explanation of these delicate effects suggest our $ab$ $initio$ GW-BSE approach is a reliable theoretical tool to interpret XAS experiments in water. 

\vspace{-1.0 em}
\section{Many-body Effects beyond the Conventional GW-BSE Calculation}
\subsection{Importance of Self-consistent Quasi-particle in Electron-hole Interaction}
\par Based on the many-body Green’s function method, the GW-BSE approach rigorously treats the electron-hole excitation via an interaction kernel~\cite{rohlfing2000,Onida2002} ($K^{\rm eh}=K^d+K^x$, or $K^{\rm eh}=K^d+2K^x$ for singlet excitations probed by photons in the absence of spin-orbit coupling) composed of both a screened Coulombic attraction $K^d$ and an unscreened exchange repulsion $K^x$. In conventional calculations~\cite{rohlfing2000,Onida2002}, the G$_0$W$_0$-BSE scheme is often adopted, in which the QP wavefunctions are approximated by the Kohn-Sham (KS)~\cite{kohn1965self} orbitals that determine the electronic ground state density within DFT. The KS orbitals give only the ground-state density, and not the electronic ground state wavefunction which can be quite different from that of the fictitious KS noninteracting system. In ordinary materials, the G$_0$W$_0$-BSE scheme is reasonable as long as the DFT XC functional gives KS orbitals that are close to the QP wavefunctions~\cite{rohlfing2000,vinson2011}. However, the above assumption needs to be carefully revisited in water, in view of the sensitivity of the H-bonds to the electron delocalization error~\cite{cohen2008i,cohen2012c}. To some extent, the inclusion of some element of the exact exchange through the use of a hybrid functional may improve the description of the KS orbitals, but it is still expected to suffer from spurious charge delocalization and intrinsic nonlocal and energy-dependent effects of the self-energy operator~\cite{zhang2021m}. In principle, the KS orbitals may still be different from the QP wavefunctions even if one has the exact XC functional and no self-interaction effects.  

\begin{figure*}[htbp]
	\setlength{\abovecaptionskip}{0.cm}
	\includegraphics[width=6.9in]{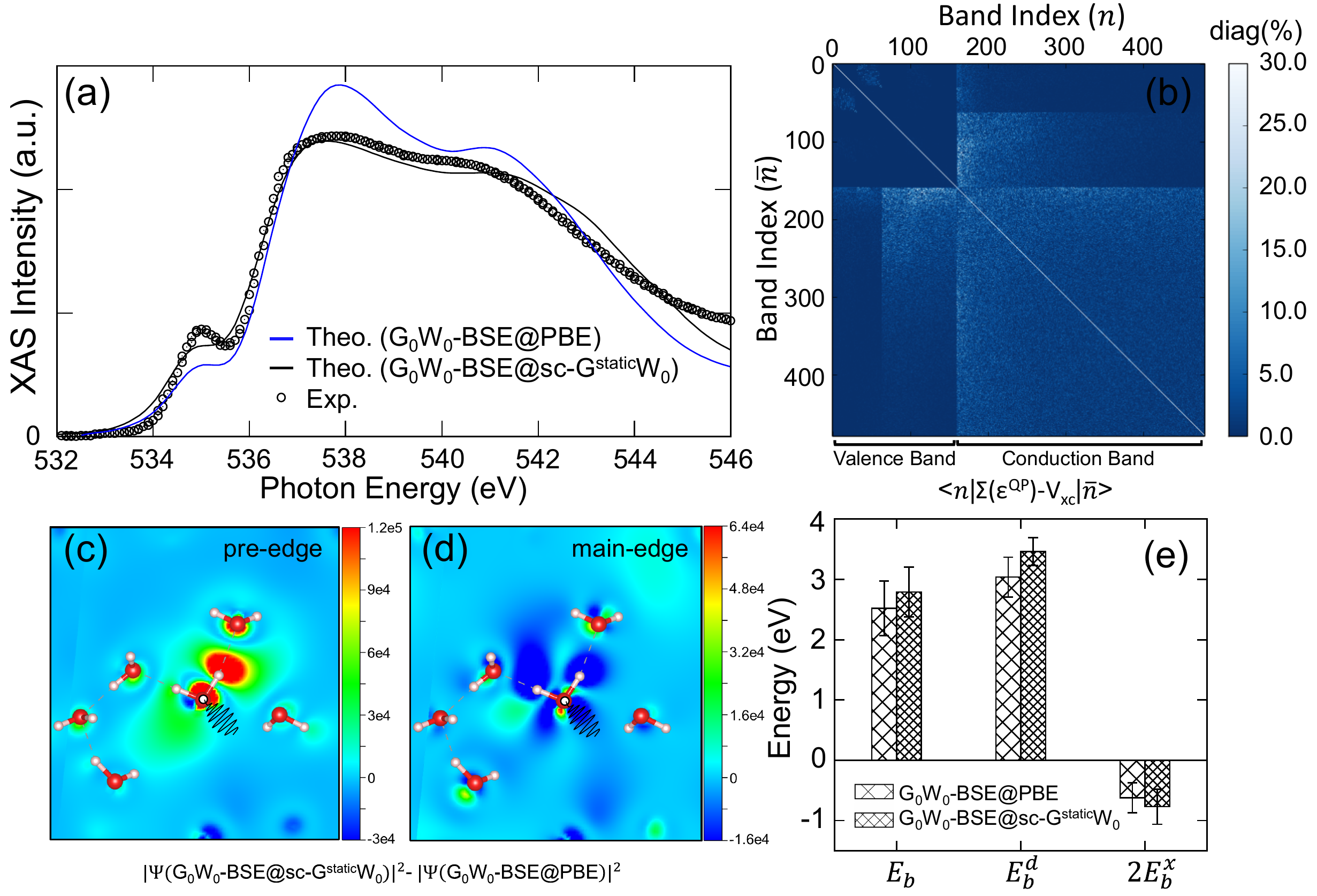}
	\caption{\label{fig:xas_g1w0}
	(a) The theoretical XAS of liquid water based on G$_0$W$_0$-BSE@PBE (blue) and G$_0$W$_0$-BSE@sc-G$^{\rm static}$W$_0$ (black) approaches. The experimental data~\cite{wernet2014} is shown with circles. (b) Visualization of the matrix elements of the static GW self energy in the KS DFT basis illustrates the difference between DFT and QP eigenstates. The matrix elements are normalized by the diagonal terms. The 2D contour plot of the electron density ($|\Psi|^2$) difference between G$_0$W$_0$-BSE@sc-G$^{\rm static}$W$_0$ and G$_0$W$_0$-BSE@PBE for excitons in the pre-edge peak (c) and main-edge peak (d) when the hole is fixed at an oxygen atom (marked with a white circle). The cutting plane is defined as the plane of H$_2$O with a hole. (e) The average exciton binding energy evaluated at the onset of the pre-edge from the G$_0$W$_0$-BSE@PBE and G$_0$W$_0$-BSE@sc-G$^{\rm static}$W$_0$ XAS, together with the contributions of direct term $E_{b}^{d}$ and exchange term $2E_{b}^{x}$ to the binding energy. }
	\vspace{-1 em}
\end{figure*}

\par In water, H-bonds are mainly a consequence of the electrostatic attraction between the positive proton and the negative lone electron pair on a neighboring water molecule. The H-bond is further stabilized by charge transfer from the occupied orbitals on the H-bond acceptor molecule to virtual orbitals on the H-bond donor molecule. Because the electrostatic contribution to the H-bond is rather weak, the charge transfer process contributes considerably to the bonding energy even though it is a higher-order many-body effect~\cite{khaliullin2008a,khaliullin2009,Misquitta2013}. This charge transfer effect should be well-described in an exact DFT. However, the practical implementations of XC potentials partially miss the discrete and quantum nature of electrons, giving rise to the well-known spurious self-interaction and lack of derivative discontinuity in DFT, which results in an overestimation of the charge transfer energy due to over-delocalized and mispositioned electronic states~\cite{caruso2012,faber2013,Kaplan2015}. Thus, unsurprisingly, in our calculations, the KS DFT orbitals are notably different from the QP wavefunctions, as evidenced by the large off-diagonal matrix elements of the static-COHSEX self energy operator in the KS DFT orbital basis, $<n|\Sigma({\varepsilon}^{\rm QP})-V_{\rm xc}|\bar{n}>$ in Fig.~\ref{fig:xas_g1w0}(b). The deviations are largest for matrix elements between the occupied $p$ bands and the unoccupied state close to Fermi level, since these are the electronic states involved in the charge transfer processes. The inaccuracy of using the KS DFT orbitals as QP wavefunctions in the G$_0$W$_0$-BSE@PBE calculation are carried over to the electron-hole excitations resulting in poor agreement with experiment in the XAS in Fig.~\ref{fig:xas_g1w0}(a). The over-delocalized KS DFT orbitals results in an underestimation of the electron-hole interactions at the BSE level. The artificially weakened electron-hole interaction in turn gives preferences to intermolecular electron-hole excitons instead of the intramolecular ones. Indeed, compared to experimental data, the XAS spectrum predicted by G$_0$W$_0$-BSE@PBE in Fig.~\ref{fig:xas_g1w0}(a) has a weaker pre-edge intensity and a more pronounced main-edge intensity. This is consistent with the fact that oscillation strengths are incorrectly transferred from intra-molecular to inter-molecular excitations.

\par In order to accurately model the excitonic effect in water, the self-consistent QP wavefunctions~\cite{Bruneval2006,Gatti2007,Vidal2010,Vidal2010b,caruso2012,Rangel2012,faber2013,Kaplan2015} should be used in describing the electron-hole interaction. Thus, we compute the XAS in water based on the G$_0$W$_0$-BSE@sc-G$^{\rm static}$W$_0$ approach, in which the QP wavefunctions are obtained by diagonalizing the single-quasi-particle Hamiltonian with the static-COHSEX self-energy. The adoption of the self-consistent QP wavefunction reduces the delocalization error of the QP orbitals near the Fermi level, and corrects the overestimated charge transfer process along the H-bond direction in DFT. Consequently, the core-hole is more attractive to the excited electrons, thus increasing the oscillator strength of the intramolecular excitons at the pre-edge. In Figs.~\ref{fig:xas_g1w0}(c) and~\ref{fig:xas_g1w0}(d), we present the difference in the electron density of the exciton wavefunction, when the position of the hole is fixed on a single oxygen atom for exciton states obtained by G$_0$W$_0$-BSE@sc-G$^{\rm static}$W$_0$ and G$_0$W$_0$-BSE@PBE approaches, for typical excitations at the pre-edge and main-edge of the XAS. For the exciton at the pre-edge in Fig.~\ref{fig:xas_g1w0}(a), it can be clearly seen that the excited electron becomes more localized around the excited water molecule, where the hole is positioned. The above largely enhances the intensities of the pre-edge feature in XAS resulting in improved agreement with experiment. Consistent with the increased localization near the core hole, the average binding energy at the onset of the pre-edge evaluated with respect to the QP transition energy is increased by 0.24 eV. Further decomposition of the binding energy (Fig.~\ref{fig:xas_g1w0}(e)) reveals that the enhanced binding energy ($E_{b}= E_{b}^{d}+2E_{b}^{x}$) is due to the increased direct Coulomb attraction $K^d$ between the oxygen core-hole and more localized excited electron state in the self-consistent QP approach. At the same time, the magnitude of the repulsive exchange interaction $K^x$ increases slightly since the excited electron and hole come closer in real space. As dictated by the optical sum rule, the increased oscillator strength of the pre-edge feature in G$_0$W$_0$-BSE@sc-G$^{\rm static}$W$_0$ shifts the oscillator strength away from the main-edge and post-edge. Therefore, an opposite picture occurs. For a typical electron-hole excitation at the main-edge, the excited electron becomes more delocalized around the excited water molecule in G$_0$W$_0$-BSE@sc-G$^{\rm static}$W$_0$ compared to that predicted by the G$_0$W$_0$-BSE@PBE scheme as shown in Fig.~\ref{fig:xas_g1w0}(d). The resulting decreased intensities of main-edge and post-edge in G$_0$W$_0$-BSE@sc-G$^{\rm static}$W$_0$ again improves agreement between the theoretical XAS and the experimental measurement.

\begin{figure}[tp]
	\setlength{\abovecaptionskip}{0.cm}
	\includegraphics[width=3.4in]{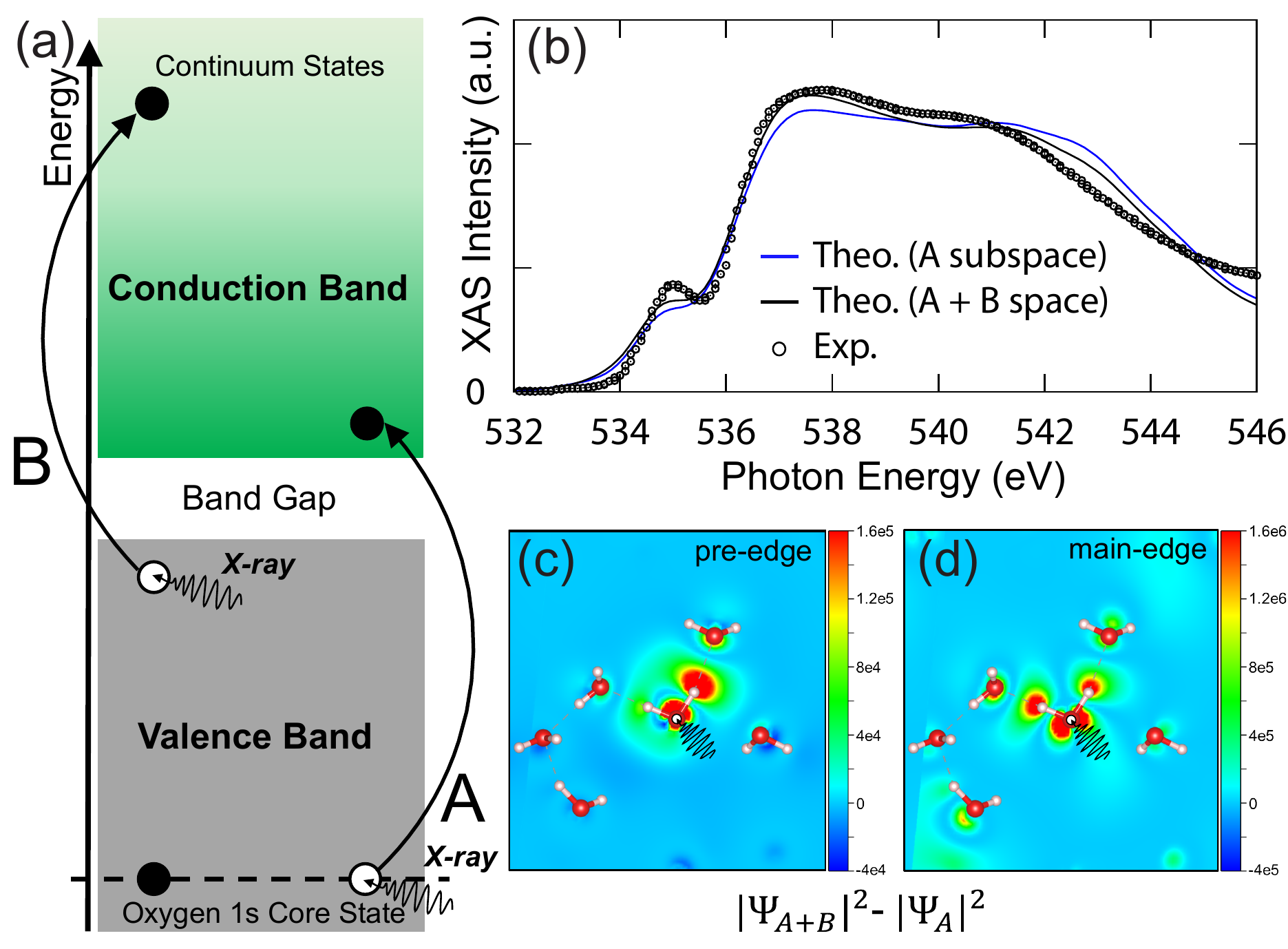}
	\caption{\label{fig:xas_screenedX}
(a) Schematic of the transitions contribute to the experimental XAS measurement, $\cal{A}$: between core states and low-lying conduction states, $\cal{B}$: between valence states and continuum states. (b) The theoretical XAS of liquid water based on the BSE kernel within $\cal{A}$ space only (blue) and within $\cal{A}\oplus \cal{B}$ space (black). The experimental data~\cite{wernet2014} is shown as circles. The 2D contour plot of the electron density ($|\Psi|^2$) difference of the exciton with the hole fix at an oxygen atom (marked with a white circle) of the pre-edge (c) and main-edge (d) peaks between GW-BSE spectra within the $\cal{A}$ subspace only and the $\cal{A}\oplus \cal{B}$ space, respectively. The cutting plane is defined as the plane of H$_2$O with a hole.}
	\vspace{-1 em}
\end{figure}
\vspace{-1 em}
\subsection{Coupling of Core-level and Valence-level Transitions in XAS}
\par Conventional theoretical treatments of XAS only consider a Hilbert space restricted to the occupied states and the lowest unoccupied (conduction) states. We will refer to this as subspace $\cal{A}$. Many-body interactions correlate electron-hole excitations in the same energy range~\cite{Fano1968}. Thus, rigorously, the theoretical treatment of core-level spectra within the BSE approach should include not only the core hole and lowest unoccupied states but also their coupling to the manifold of all other transitions with energy in the range of the core-level excitation (subspace $\cal{B}$). A proper description of XAS thus demands the modeling of all possible transitions between the initial and final states within the energy of the X-ray photon. As schematically shown in Fig.~\ref{fig:xas_screenedX}(a), the complete Hilbert space includes not only transitions between the oxygen 1s core state and the excited electrons in low energy regions of conduction band as denoted by subspace $\cal{A}$, but also transitions in the subspace $\cal{B}$, between the states in the valence band states of water and electrons that are excited to the continuum states in conduction band at much higher energies corresponding to the energy of the X-ray photon~\cite{PARRATT1959}. The calculation of XAS of water within the entire Hilbert space ($\cal{H} = \cal{A} \oplus \cal{B}$) encounters significant barrier in computing tens of thousands of QP wavefunctions, and the subsequent treatment of a large number of electron-hole pair interactions in the BSE is computationally unfeasible within the constraint of current computer resources. Therefore, all theoretical calculations to date~\cite{Rehr2000,fransson2016x} have considered only electron-hole pairs within the subspace $\cal{A}$, and it was assumed that couplings to the continuum of transitions from subspace $\cal{B}$ are small.

\par Here, we employ a new method to account for the coupling between core-level excitons formed within subspace $\cal{A}$ and the continuum of valence-level transitions through a matrix downfolding approach~\cite{DianaTBD}. In this approach, the full Hilbert space is downfolded to subspace $\cal{A}$. For core excitons, the contribution of the direct Coulomb interaction to the downfolding is small due to the orthogonality of core states in subspace $\cal{A}$ and valence states in subspace $\cal{B}$. For the exchange interaction, the downfolding is exactly equivalent to screening the exchange term with the polarizability due to electron-hole transitions in subspace $\cal{B}$~\cite{benedict2002,Deilmann2019,diana2021,DianaTBD}. Using this approach allows us to compute the XAS of liquid water including the coupling of the core-conduction and valence-conduction transitions. Even within this approach, the calculation of the dielectric screening for the downfolding remains a significant computational cost when extended to multiple snapshots of liquid water. Consequently, we perform one fully \textit{ab initio} calculation of the screened exchange and use that to parametrize a scaling of the exchange interaction that mimics the \textit{ab initio} screening. (see section II.C)

\par The resulting XAS within the subspace $\cal{A}$ only and the downfolded Hilbert space ($\cal{H} = \cal{A} \oplus \cal{B}$) are shown in Fig.~\ref{fig:xas_screenedX}(b). It can be clearly seen in Fig.~\ref{fig:xas_screenedX}(b) that the spectrum generated from the downfolding of the complete Hilbert space is well-approximated by the spectrum from subspace $\cal{A}$. This is consistent with the expectation that the strong excitonic effects results in a large oscillator strength for the core-level excitons and comparatively weak coupling to the valence continuum. Nevertheless, the transition from subspace $\cal{B}$ provides a small but noticeable renormalization effect to the overall XAS. Within the downfolding approach, the renormalization effects from subspace $\cal{B}$ are equivalent to applying a screened exchange $K^x_{\rm eff}$ on the coupling of the electron-hole excitations in subspace $\cal{A}$. The less repulsive electron-hole interaction means that the overall electron-hole interaction becomes more attractive via the GW-BSE kernel ($K^{\rm eh}=K^{d}+K^x_{\rm eff}$). In particular, the excitons at the pre-edge and main-edge respectively become even more strongly bound around the core-hole as shown in Fig.~\ref{fig:xas_screenedX}(c) and ~\ref{fig:xas_screenedX}(d), which unsurprisingly, yields enhanced oscillator strength at both the pre-edge and the main-edge. With the above renormalization, the intensities at post-edge slightly decreases as dictated by the optical sum rule. Compared to the conventional XAS computed from subspace $\cal{A}$ only, the current theoretical XAS in water considering the full Hilbert space yields a noticeably improved agreement with experiment.

\vspace{-1.0 em}
\section{conclusions}
\par  In conclusion, we show that the XAS of liquid water can be accurately predicted by electron-hole excitation theory through the $ab$ $initio$ GW-BSE approach. In addition to the self-energy and excitonic effects, more sophisticated many-body effects beyond conventional one-shot GW-BSE approaches are required for calculations on water due to the delicate nature of the H-bond network. On the one hand, QP wavefunctions, obtained by diagonalizing the self-energy operator, should be used instead of the conventional KS DFT orbitals. The employment of self-consistent QP wavefunctions corrects the overestimated charge transfer processes in DFT and yields significantly improved spectral features. On the other hand, we have also considered electron-hole excitations in the complete Hilbert space, which further yields a noticeable improvement in the agreement of the XAS with experiment. The previously neglected interactons between core-level excitations and the excitations of valence band to higher-energy conduction band results in a non-negligible renormalization of the XAS. Interestingly, the agreement of the calculated spectra with experiment showed systematic improvement with the removal of key simplifying approximations.  
\par Our current work provides a rigorous framework for solving the long-standing challenge of modeling the XAS of liquid water from first-principles. The accurate predictions from our approach will further help to resolve the controversy of the last two decades in the spectral interpretation of the underlying H-bond network. Our simulations show that a fully first-principles approach can lead to XAS spectra in good agreement with experiment without requiring a drastic revision of the standard picture of the tetrahedral H-bond network of water. Furthermore, our work suggests that the high-level many-body effects can play crucial roles in the electron-hole excitations for materials such as water. With further algorithmic optimization and access to advanced computer platforms, the methodologies developed in this work are ready to be widely applied to aqueous systems, such as ionic solutions, confined water, water at interfaces, as well as other donor-acceptor systems, such as molecular crystals, where the charge-transfer effect is important.

\vspace{-1.0 em}
\section*{acknowledgments}

\par This work was primarily supported by the Computational Chemical Center: Chemistry in Solution and at Interfaces funded by the DoE under Award No. DE-SC0019394 as well as the Computational Materials Science Center:  Center for Computational Study of Excited-State Phenomena in Energy Materials funded by the DoE under Contract No. DE-AC02-05CH11231. The work of F. T., C. Z., R. C. and X. W. of MD simulation and XAS calculation was supported by the Computational Chemical Center: Chemistry in Solution and at Interfaces funded by the DoE under Award No. DE-SC0019394. The work of Z. L., S. G. L., and D. Y. Q. related to implementation of new methodology and algorithms to calculate XAS with the BerkeleyGW code were supported by the Center for Computational Study of Excited-State Phenomena in Energy Materials at the Lawrence Berkeley National Laboratory funded by the DOE under Contract No. DE-AC02-05CH11231, as part of the Computational Materials Sciences Program. The work of D. Y. Q. on the development of the Hilbert space downfolding approach for core-level spectroscopy was supported by the National Science Foundation (NSF) under grant number DMR-2114081. The work by C.Z. on water structure by neural network potential was supported by National Science Foundation through Awards DMR-2053195. This research used resources of the National Energy Research Scientific Computing Center (NERSC), which is supported by the U.S. Department of Energy (DOE), Office of Science under Contract No. DE-AC02-05CH11231. This research used resources of the Oak Ridge Leadership Computing Facility at the Oak Ridge National Laboratory, which is supported by the Office of Science of the U.S. Department of Energy under Contract No. DE-AC05-00OR22725. This research includes calculations carried out on HPC resources supported in part by the National Science Foundation through major research instrumentation grant number 1625061 and by the US Army Research Laboratory under contract number W911NF-16-2-0189.


\bibliography{bse}

\end{document}